\documentclass[useAMS,usenatbib]{mn2e}

\usepackage{url}
\usepackage{lscape}
\usepackage{graphicx}
\usepackage{float}
\usepackage{upgreek}
\usepackage{floatflt}

\title{A search for optical bursts from the repeating fast radio burst FRB 121102}
\author{L. K. Hardy,$^{1}$ V. S. Dhillon,$^{1,2}$ L. G. Spitler,$^{3}$ S. P. Littlefair,$^{1}$ R. P. Ashley,$^{4}$ 
\newauthor
A. De Cia,$^{5}$ M. J. Green,$^{4}$ P. Jaroenjittichai,$^{6}$ E. F. Keane,$^{7,8}$ P. Kerry,$^{1}$  
\newauthor
M. Kramer,$^{3,8}$ D. Malesani,$^{9}$ T. R. Marsh,$^{4}$ S. G. Parsons,$^{1}$ A. Possenti,$^{10}$ 
\newauthor
S. Rattanasoon$^{1,6}$ and D. I. Sahman$^{1}$\\
$^{1}$Department of Physics and Astronomy, University of Sheffield, Sheffield, S3 7RH, UK\\
$^{2}$Instituto de Astrof\`{i}sica de Canarias, E-38205 La Laguna, Tenerife, Spain\\
$^{3}$Max-Planck-Institut f\"{u}r Radioastronomie, Auf dem H\"{u}gel 69, D-53121 Bonn, Germany\\
$^{4}$Department of Physics, University of Warwick, Coventry CV4 7AL, UK\\
$^{5}$European Southern Observatory, Karl-Schwarzschild Str. 2, 85748 Garching bei M\"{u}nchen, Germany\\
$^{6}$National Astronomical Research Institute of Thailand, 191 Siriphanich Building, Huay Kaew Road, Chiang Mai 50200, Thailand\\
$^{7}$Square Kilometre Array Organisation, Jodrell Bank Observatory, Macclesfield SK11 9DL, UK\\
$^{8}$Jodrell Bank Centre for Astrophysics, School of Physics and Astronomy, University of Manchester, Manchester M13 9PL, UK\\
$^{9}$Dark Cosmology Centre, Niels Bohr Institutet, University of Copenhagen, Juliane Maries Vej 30, DK-2100 Copenhagen, Denmark\\
$^{10}$Instituto Nazionale di Astrofisica (INAF) – Osservatorio Astronomico di Cagliari, Via della Scienza 5, I-09047 Selargius (CA), Italy}

\begin{document}

\date{}

\pagerange{\pageref{firstpage}--\pageref{lastpage}} \pubyear{2017}

\maketitle

\label{firstpage}

\begin{abstract}
We present a search for optical bursts from the repeating fast radio burst FRB\,121102 using simultaneous observations with the high-speed optical camera ULTRASPEC on the 2.4-m Thai National Telescope and radio observations with the 100-m Effelsberg Radio Telescope. A total of 13 radio bursts were detected, but we found no evidence for corresponding optical bursts in our 70.7-ms frames. The 5-$\sigma$ upper limit to the optical flux density during our observations is 0.33 mJy at 767nm. This gives an upper limit for the optical burst fluence of $0.046$ Jy~ms, which constrains the broadband spectral index of the burst emission to $\alpha \le -0.2$. Two of the radio pulses are separated by just $34$\,ms, which may represent an upper limit on a possible underlying periodicity (a rotation period typical of pulsars), or these pulses may have come from a single emission window that is a small fraction of a possible period.

\end{abstract}



\section{Introduction}
Fast radio bursts (FRBs) are short, bright pulses of coherent radio emission, with dispersion measures (DMs) in excess of that expected for Galactic sources. Since the first discovery by \citet{lorimer07}, a further 22 FRBs have been reported in the FRB catalogue\footnote{\url{http://www.astronomy.swin.edu.au/pulsar/frbcat/}} \citep{petroff16}. The high DMs (a measure of the frequency-dependent arrival times of the radio pulses, due to propagation through a low density plasma) lead us to infer extragalactic distances to FRBs. This makes them potentially valuable cosmological probes which could provide a direct measurement of the cosmic density of ionized baryons in the intergalactic medium, probe intergalactic magnetic fields, and constrain the dark energy equation of state parameter (see e.g. \citealp{zhou14,keane16,ravi16}, and references therein). There appear to be almost as many proposed progenitor theories for FRBs as burst detections themselves. Many of these are either ruled out or not yet tested, but the leading theories include supergiant radio pulses, magnetar giant flares and the collapse of supramassive neutron stars in the case of non-repeating events (see e.g. \citealp{katz16b,katz16} for reviews). 

FRB\,121102 is the first and only FRB to show repeated bursts at a consistent sky position and DM \citep{spitler14,spitler16}. The observed bursts have DMs consistent with being constant; the mean value is $559.6$ pc cm$^{-3}$ with a 1-$\sigma$ RMS of 4.2 pc cm$^{-3}$ and the mean of the quoted uncertainties is 4.8 pc cm$^{-3}$ \citep{spitler16,scholz16}. FRB\,121102 has recently been localised to a faint dwarf galaxy at redshift $z=0.19$ \citep{chatterjee17,marcote17,tendulkar17}. The extragalactic localisation and the repeating nature of FRB\,121102 enables the rejection of certain progenitor scenarios involving catastrophic events that do not repeat \citep{totani13,kashiyama13,falcke14,zhang16b}, for this particular burst at least. The repeating nature may favour a young magnetar scenario \citep{metzger17}, a pulsar emitting giant pulses \citep{cordes16}, or may be triggered by an AGN \citep{katz17,vieyro17}. However, it is unclear whether FRB\,121102 may be fundamentally different to the other known FRBs, none of which have been seen to repeat. The recent observation that FRB\,121102 appears to be coincident with a star forming region in its host galaxy \citep{bassa17} lends support to the possible neutron star scenarios \citep{lyutikov16,katz16c}. The pulses from FRB\,121102 have intrinsic durations of at most several milliseconds, and this time scale may be an additional indication of an underlying neutron star explanation. However other FRBs are seen to be unresolved in time, even with a resolution limit of a few 100 $\upmu$s. This could hint at a divergence in the progenitors of the FRB population \citep{scholz16}.

\begin{table*}
\begin{minipage}{1.\textwidth}
\centering\footnotesize
\caption{Observations conducted with ULTRASPEC on the TNT. The $i'+z'$ filter is a custom-made broad band filter comprising the SDSS $i'$ and $z'$ passbands}
\label{tab:obs}
\begin{tabular}{p{1.5cm} p{1.6cm} p{1.6cm} p{1.7cm} p{1cm} p{2.6cm} p{5cm}}
\textbf{Date} & \textbf{Start time (UTC)} & \textbf{End time (UTC)} & \textbf{t$\mathrm{_{exp}}$ (ms)} & \textbf{Filter} & \textbf{No. of coincident radio bursts} & \textbf{Conditions}  \\ \hline
2017-01-16 & 15:50:00 & 19:45:12 & 70.7 & $i'\!+\!z'$ & 3 & Clear, $3''$ seeing \\
2017-01-19 & 15:35:14 & 18:46:45 & 70.7 & $i'\!+\!z'$ & 5 & Clear, $1''$ seeing \\
2017-01-25 & 13:29:54 & 14:31:15 & 70.7 & $i'\!+\!z'$ & 0 & Clear, $3''$ seeing \\
2017-01-25 & 15:15:20 & 19:00:22 & 70.7 & $i'\!+\!z'$ & 3 & Clear, $3''$ seeing \\
2017-02-16 & 12:43:21 & 17:32:27 & 70.7 & $i'$ & 0 & Intermittent clouds, $1''$ seeing \\
2017-02-19 & 14:34:38 & 17:32:24 & 70.7 & $i'$ & 1 & Intermittent clouds, $1.5''$ seeing \\
 \hline  
\end{tabular}
\end{minipage}
\end{table*}

In order to constrain the emission mechanism for FRB\,121102, we need to investigate the burst behaviour at other wavelengths. For example, if the FRB is powered by a neutron star, in certain scenarios one might expect to see an optical burst counterpart. The pulsar supergiant pulse mechanism, as seen in the Crab pulsar \citep{lyutikov16}, would have some simultaneous optical brightening associated with the radio bursts \citep{shearer03}. Equally, there may be optical emission from magnetar giant flares \citep{lyutikov16b}. In fact, rapid optical flaring behaviour has been seen before in a gamma-ray transient with a light curve which is phenomenologically similar to gamma-ray flares from magnetars \citep{stefanescu08}. An optical burst detection may also provide an insight into the local environment of the burst engine \citep{metzger17}. 

Our investigation follows the same technique employed by \citet{dhillon11} in their search for optical pulses from Galactic rotating radio transients, as this technique allows for a much deeper search for optical counterparts than any single long-exposure image (see also the discussion by \citealp{lyutikov16b}). If any optical burst emission exists, we expect it to be short-lived, and thus we require fast observations to probe any variability which might be present. Equally, as we are interested in the actual energy release of the emission mechanism, we need to measure the fluence of the source, rather than the flux density, and this can be done more accurately with shorter exposures. 

More importantly, if the burst signal is very weak the signal-to-noise ratio (S/N) will be dominated by the shot noise from the sky background, which increases with the length of exposure. Consider a burst which emits 100 photons in a few milliseconds, and a 10-ms optical exposure which collects those 100 burst photons as well as $N$ sky photons. Meanwhile, a 1000-ms exposure would contain the same 100 burst photons, but this time it would also record $100N$ sky photons. The shot noise contribution from the sky will be $\sqrt{100}=10$ times larger in the longer exposure, and thus the signal-to-noise ratio will be lower. In fact it is possible that any short-lived optical signal may be swamped by the shot noise in the sky background, which is inevitable in deep images. The search for an optical burst counterpart to FRBs thus requires high-speed imaging, because it allows us to probe fainter fluence limits on the detection of optical emission. 


\section{Observations}\label{sec:obs}
\subsection{Optical Observations with ULTRASPEC}
We conducted high-speed optical imaging of the field of FRB\,121102 during January and February 2017 with ULTRASPEC, mounted on the 2.4-m Thai National Telescope (TNT) at Doi Inthanon \citep{dhillon14}. ULTRASPEC is purpose built for high-speed imaging, with an electron-multiplying frame-transfer CCD and a data acquisition system optimised for fast readout with essentially no read-out noise \citep{tulloch11}. We present a total of 19.6 hours of high-speed photometry, during which time we also detected 12 radio bursts with the 100-m Effelsberg Radio Telescope (see Section \ref{sec:effelsberg}). This allowed us to pinpoint the exact images in which we might expect an optical burst counterpart, if the optical and radio bursts are emitted by the source simultaneously. Journals of the optical and radio observations are presented in Tables \ref{tab:obs} and \ref{tab:radioObs}.

\begin{figure}
 \includegraphics[width=.49\textwidth,angle=270]{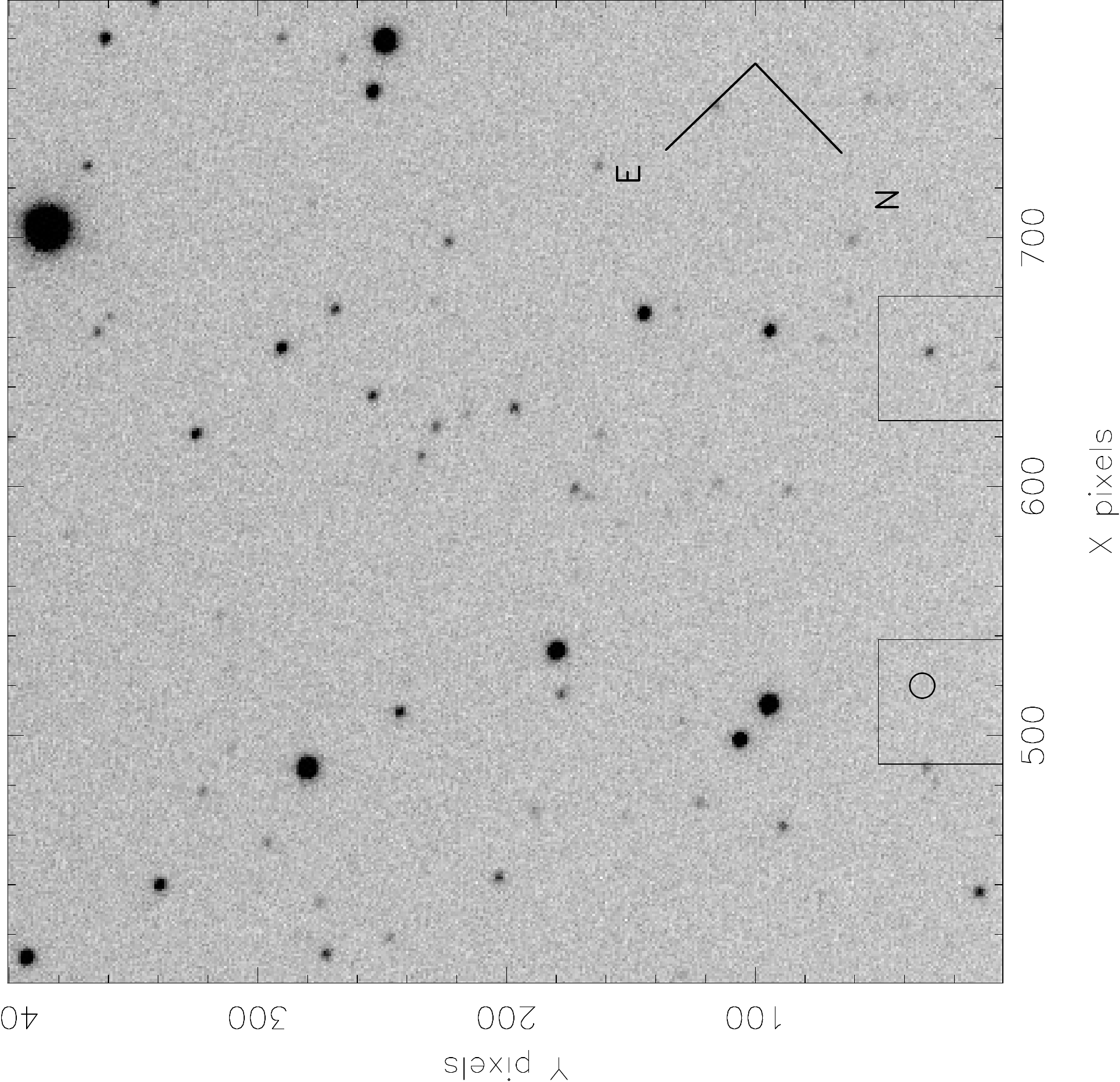}
 \caption{Finding chart showing the location of the two ULTRASPEC drift-mode windows used, which are 46 by 50 unbinned pixels, covering an area of $20.8''$ by $22.6''$. The location of FRB\,121102 is shown with the small circle in the left-hand window. The comparison star in the right-hand window is PSO J053154.854+330815.385.}
 \label{fig:finder}
\end{figure}

Assuming that FRBs are powered by a neutron star, we can make a rough estimate of the expected optical burst duration. The Crab pulsar shows optical pulses which are up to five times wider than the associated giant radio pulses, and which lead the radio pulses by up to 0.4ms \citep{shearer03,slowikowska09}. Assuming that the FRB mechanism follows similar behaviour to the Crab pulsar, we might expect the optical and radio bursts to be effectively coincident in time, and that the radio bursts are equal or shorter in duration than any optical bursts. The radio bursts have durations of $\sim$2--7 ms \citep{scholz16}, and thus we might expect that any optical bursts would have a duration of less than $\sim$10--35~ms. 

ULTRASPEC was therefore operated in its fastest windowed `drift' mode (see \citealp{dhillon07b}), using two drift-mode windows each of size 50 by 46 unbinned pixels, with on-chip binning by a factor of two, such that the binned pixels had a plate scale of $0.9\,''\!$/pixel. This allowed us to reach a frame rate of $\sim14$ Hz with a deadtime of only 0.7~ms (a duty cycle of 99\%). We used the electron-multiplying (EM) read-out mode, which has an effective readout noise of 0.005 electrons, and an effective electron-multiplying gain (the number of measured photo-electrons for each input electron) of the order of 1200. Throughout this paper discussion of measured counts are quoted in EM-boosted ADU, and have not been scaled to original photo-electrons. One window was placed on the location of the FRB found by \citet{marcote17}, with RA 05:31:58.70, DEC +33:08:52.5 (J2000), whilst the other window covered a nearby comparison star, PSO J053154.854+330815.385. Figure \ref{fig:finder} shows the on-sky location of the two windows. It is possible to run ULTRASPEC faster than the 70-ms cadence used in these observations, but this requires smaller windows and the TNT has no autoguider: we did not want to risk having the FRB location and comparison star drift outside of the windows. A total of 995\,900 images were obtained.

\subsection{Radio Observations with Effelsberg}\label{sec:effelsberg}
In addition to the optical observations, we conducted simultaneous radio coverage of FRB\,121102 with the 100-m radio telescope at Effelsberg, Germany. We detected 13 distinct radio bursts, though one of these was outside the time-window of the optical observations with ULTRASPEC. The radio bursts were detected using the 7-beam feed array, and the receiver has a system equivalent flux density of 17 Jy and a full-width at half-maximum beam size of 9.8 arcmin. The central pixel was pointed to the sky position of FRB\,121102 as found by \citet{chatterjee17}. The observing configuration was identical to that used by the High Time Resolution Universe-North (HTRU-N) pulsar and fast transient survey for which details can be found in \citet{barr13}. 

High time-resolution total-intensity spectral data were recorded with the Pulsar Fast-Fourier-Transform Spectrometer (PFFTS) search mode backends from all seven beams, although we only expected bursts from the central beam. The backends provide a bandwidth of 300 MHz centered at 1360 MHz divided into 512 frequency channels at a time resolution of 54.613 $\upmu$s. Data from only the central beam were also recorded using the PSRIX pulsar timing backend \citep{lazarus16} in parallel, which provides 256 MHz of bandwidth centered on 1358.9 MHz divided into 256 frequency channels. The PFFTS spectrometers have better sensitivity to bursts but provide no absolute timing information, which is needed to compare the optical and radio times-of-arrival (TOAs). As described in detail below we use bursts detected with both backends to transfer the absolute timing from the PSRIX to the PFFTS detections as needed.

\begin{table}
\centering\footnotesize
\caption{Observations conducted with the Effelsberg 100m radio telescope.}
\label{tab:radioObs}
\begin{tabular}{p{1.45cm} p{1.75cm} p{1.75cm} p{1.9cm}}
\textbf{Date} & \textbf{Start time (UTC)} & \textbf{End time (UTC)} & \textbf{No. of bursts observed} \\ \hline
2017-01-16 & 16:14:20 & 20:14:20 & 3 \\
2017-01-19 & 16:05:40 & 19:36:10 & 5 \\
2017-01-25 & 16:22:00 & 19:52:00 & 4 \\
2017-02-16 & 14:09:40 & 17:46:58 & 0 \\
2017-02-19 & 14:28:40 & 17:28:40 & 1 \\
 \hline  
\end{tabular}
\end{table}

\begin{figure*}
 \includegraphics[width=1.22\textwidth,angle=270]{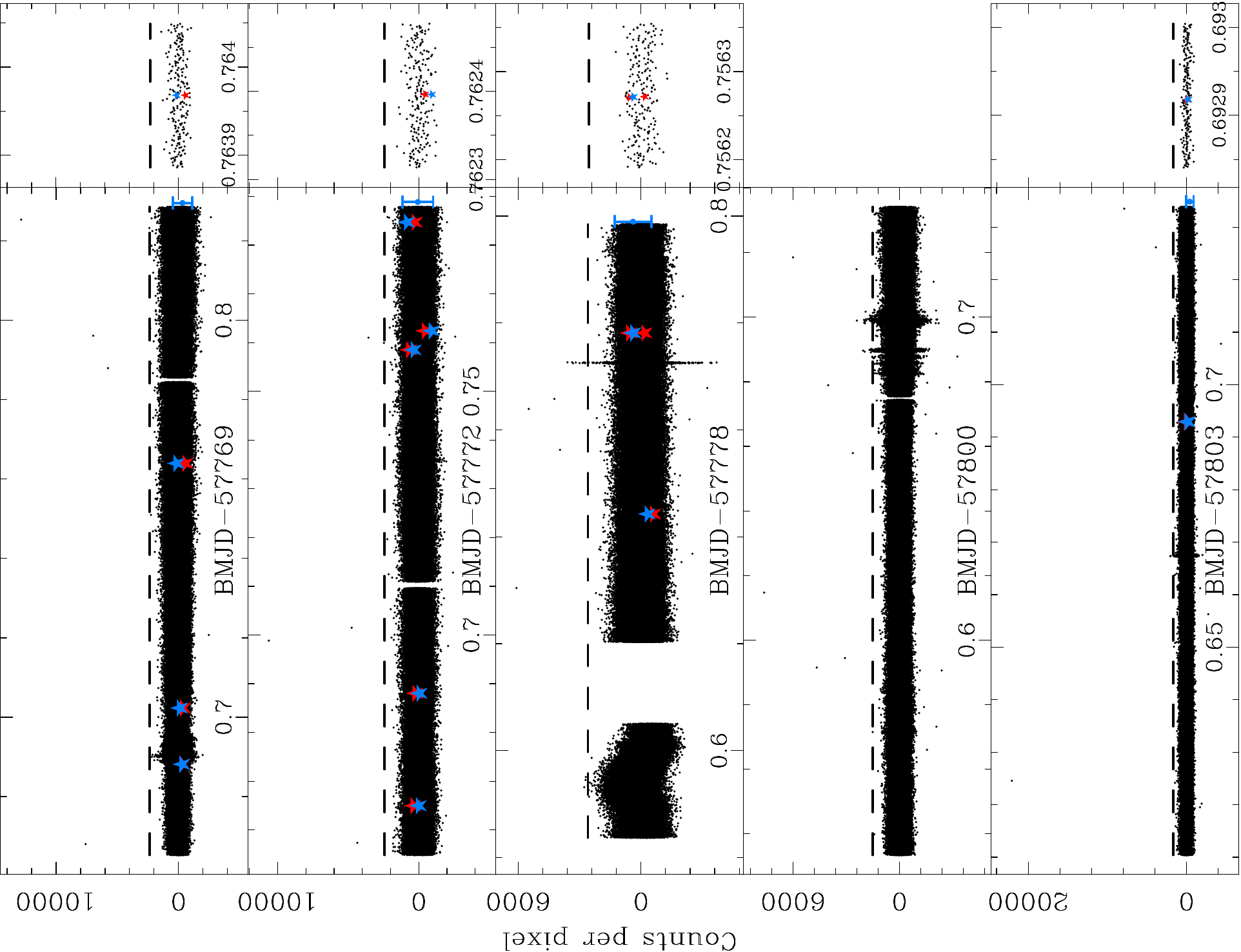}
 \caption{ULTRASPEC light curves of FRB\,121102 showing the sky-subtracted counts per pixel as a function of time. The sub-plots on the right-hand side show a zoomed-in view of the light curves close to the arrival times of a sample of the Effelsberg radio burst detections. The occasional short periods of larger scatter are likely due to the increase in sky noise during passing cloud cover. The dashed line shows the 5-$\sigma$ level for each light curve, where $\sigma$ is the standard deviation in the counts. Red stars show the single ULTRASPEC measurements for the frames closest in time to the simultaneous Effelsberg radio burst detections. Blue stars represent the summed optical flux (counts/pixel) from the two temporally closest frames to a radio burst detection, in case the burst may be spread across two frames. The single blue data point at the right-hand side of each plot represents the stacked flux (counts/pixel) from all bursts in each run. The error bar is the 1-$\sigma$ error in this value. Two bursts (the second and third observed in the third observing run) were observed just 34~ms apart.}
 \label{fig:lightcurves}
\end{figure*}

\section{Analysis}\label{sec:analysis}
We performed aperture photometry on the optical data at the expected position of FRB\,121102, using fixed-size circular apertures of radius $2.7''$. This aperture size was chosen to accommodate the positional uncertainty of the FRB location and the average seeing throughout the observations, and was kept fixed in an attempt to keep the sky-noise contribution as stable as possible throughout the analysis. We calculated the expected position of the FRB using an offset from the measured position of the comparison star in the other window (see Figure \ref{fig:finder}). Offsets were calculated using a full-frame finder image with astrometric calibration performed by \url{astrometry.net} \citep{lang10} to an astrometric accuracy of $1''$. All frame times were corrected to barycentric modified Julian date (BMJD). The sky background was subtracted using the median level in an annulus around the target aperture.

The radio data were searched for single pulses using the PRESTO\footnote{\url{https://github.com/scottransom/presto}} software package \citep{ransom01}. First, an RFI mask was generated using {\tt rfifind}. The data were then dedispersed with 20 trial DMs from 550~pc~cm$^{-2}$ to 570~pc~cm$^{-2}$, i.e.\ centred on the true DM of the FRB ($\sim$560~pc~cm$^{-2}$), in steps of 2~pc~cm$^{-2}$. The data were also downsampled in time by a factor of 16, because the inter-channel dispersion smearing is $\sim$1~ms. Bursts were identified with {\tt single\_pulse\_search.py} within PRESTO \citep{lazarus15}, which convolves each dedispersed time series with a series of boxcar templates to optimize a burst's S/N and then applies a detection threshold. All candidate events with S/N $\sim7$ were inspected manually. 

Recent observations of FRB\,121102 suggest that the optimal DM for aligning burst sub-structure is DM = 560~pc~cm$^{-2}$ \citep{scholz16}, and for the remainder of the analysis we fixed the DM at this value. The TOAs were calculated from the burst arrival time relative to the start of the observation. Because only 10 of the 13 bursts were detected in the PSRIX data, we used the PFFTS detections for this analysis, but the PFFTS has no absolute timing information. Therefore, we calculated a time correction for each scan using bursts detected in both the PFFTS and PSRIX data. These topocentric TOAs (referenced to the highest frequency in the band) were converted to barycentric TOAs at infinite frequency using PRESTO's {\tt bary} function. The uncertainty on the TOAs is given by the positional uncertainty on the burst fitting described below. The TOAs are listed in Table \ref{tab:radioBursts}.

To determine the radio flux density and pulse width, a short segment of time-frequency data around each burst was extracted using {\tt dspsr} \citep{vanstraten11}. The data were averaged in frequency to generate a burst profile, which was then downsampled to a time resolution of $\sim$0.5~ms. Each burst profile was fit with a Gaussian function using a least squares routine. The radio flux densities were estimated using the radiometer equation \citep{lorimerbook} for two summed polarizations, a bandwidth of 250~MHz (the effective bandwidth of the PFFTS data after RFI flagging and removing the edges of the passband), and the maximum S/N and pulse width determined from the Gaussian fitting for each burst. The fitting results are given in Table \ref{tab:radioBursts}.

The uncertainties on the pulse width are the 1-$\sigma$ uncertainties from the least-squares fitting. The uncertainties on the radio flux density are dominated by two systematic contributions. The first is the uncertainty on the determination of the S/N, i.e. the RMS noise level, and is $\sim$15 mJy, assuming a median pulse width of 2.6~ms. The second factor is the variation in the receiver system equivalent flux density with time, which we estimate to be $\sim$15\% of the mean value. The reported mean flux density uncertainty is the RMS sum of these two factors.

\section{Results}\label{sec:results}

Figure \ref{fig:lightcurves} shows the optical light curves for the aperture centred on FRB\,121102 on the five nights of observations. With upwards of 100\,000 data points in each light curve, we require a detection at the 5-$\sigma$ confidence level, such that less than one point in $10^6$ might be found above this level in a normal distribution. Figure \ref{fig:lightcurves} shows several points above the dashed lines representing 5-$\sigma$, where $\sigma$ is the error in the optical flux (counts/pixel), and is measured as the standard deviation of each light curve. The images which correspond to the points above 5-$\sigma$ were all investigated by eye, and all apparent signals appear to be due to cosmic ray hits, judging by the size and shape of their point spread functions. 

The expected flux of cosmic rays detectable at the altitude of the TNT (2500m) by a CCD is approximately 250 m$^{-2}$s$^{-1}$ \citep{smith02spie,groom04}, although radioactive decay of compounds which make up the glass lenses and other components of instruments can increase this value by factors of up to 5 to 10 \citep{florentin95,groom04}. Nevertheless, the observed rate of cosmic ray incidence in our observations is around 1 event per hour, or 10000 m$^{-2}$s$^{-1}$, a factor of 40 times higher than expected. Such a discrepancy requires there to be a large amount of background radiation at the TNT. We conducted aperture photometry on a blank region of the field to investigate whether this observed cosmic ray rate is indeed typical, and we found a consistent hit rate. This reassures us that the points above the 5-$\sigma$ lines in Figure \ref{fig:lightcurves} are indeed false positives, despite the unusually high apparent cosmic ray flux.

In Figure \ref{fig:lightcurves} we highlight in red the measurements from the frames closest in time to the infinite-frequency arrival times of the radio bursts. None of these frames show any significant optical signal. In case a burst may be spread across two consecutive ULTRASPEC frames, we also analyse the stacked fluxes (counts/pixel) of the two temporally closest frames to the radio burst arrival times. The results of these stacked fluxes are shown in blue in Figure \ref{fig:lightcurves}: again no significant optical signals are seen. Note that two of the radio bursts observed on 2017-01-25 are so close together in time that the two nearest optical frames are the same. See Section \ref{sec:period} for further discussion on these narrowly separated bursts.

\begin{figure*}
\includegraphics[width=.49\textwidth,angle=270]{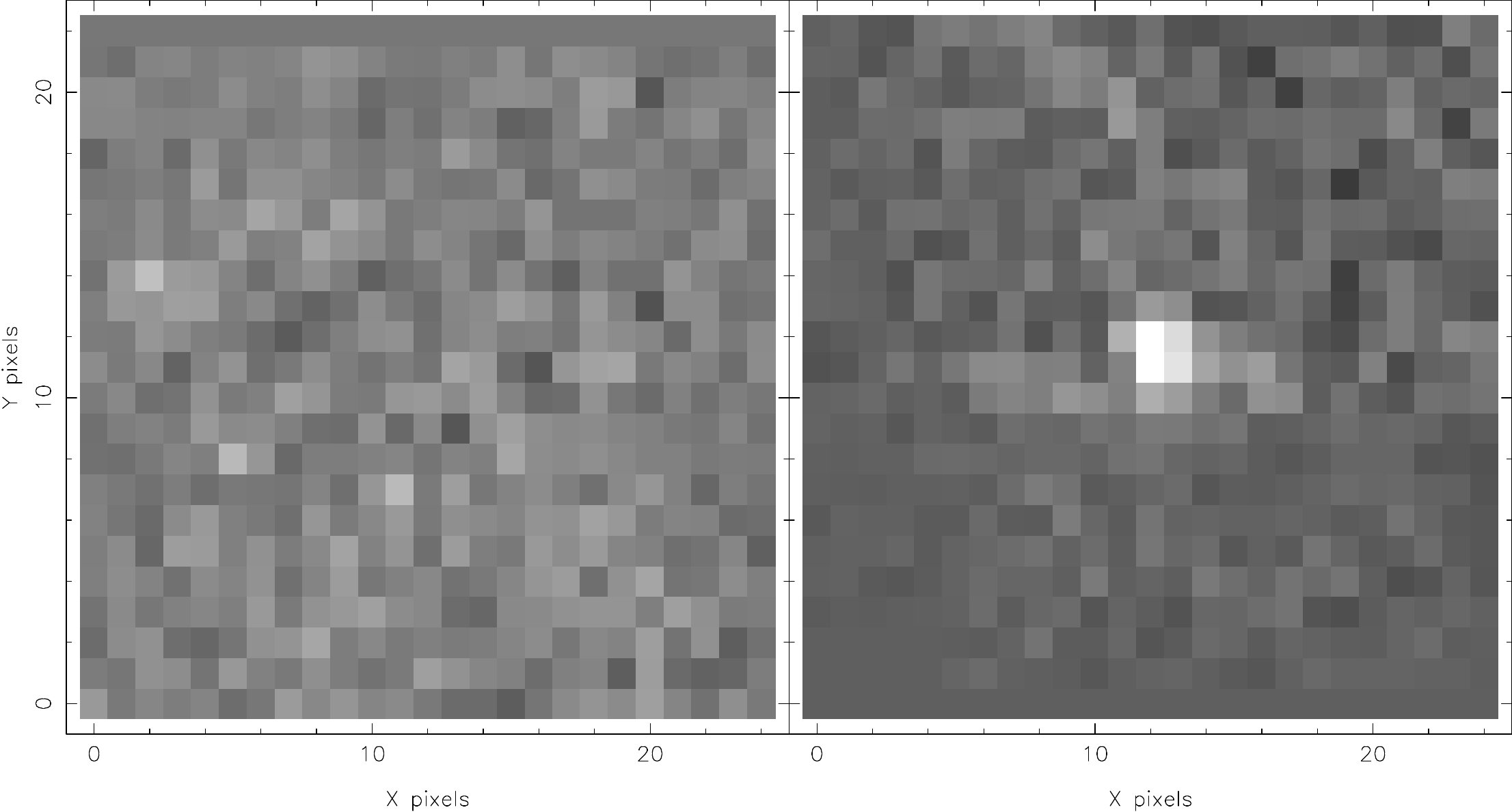}
\caption{Stacked images consisting of 22 ULTRASPEC frames observed at the same time as the 12 Effelsberg radio bursts. The left image is centred on the expected position of FRB\,121102, while the right image is centred on the comparison star PSO J053154.854+330815.385. The FWHM of the comparison star radial profile is approximately 1 binned pixel ($0.9''$). The field of view of each frame is $23''\times21''$. The background signal in the left image is $208000\pm7000$ counts per pixel. Both images have the same greyscale range.}
\label{fig:stacked}
\end{figure*}

In Figure \ref{fig:stacked} we show the result of stacking 22 frames, centred on the FRB location and corrected for tracking errors. These 22 frames include the two nearest frames to each of the 12 observed radio burst arrival times. This stacked image shows no sign of any optical signal. We also shifted and stacked all 995\,900 frames, but found no signal from the FRB or host galaxy (as expected -- we predict a S/N for the host galaxy of $\sim1$ without accounting for additional clock-induced charge, \citealp{dhillon14}).

\begin{table*}
\begin{minipage}{1.\textwidth}
\centering\footnotesize
\caption{Details of the radio bursts detected by Effelsberg. TOA is the time of arrival of the burst, scaled to infinite frequency. The width and strength of each burst are the width that maximises the S/N and the corresponding flux density value. Uncertainties in the TOA and pulse width measurements are given in brackets and represent the last one or two digits. Note that these measured pulse widths are an RMS sum of the intrinsic width and instrumental broadening due to interchannel DM smearing, which is approximately 1~ms.}
\label{tab:radioBursts}
\begin{tabular}{p{3.0cm} p{2.7cm} p{3.0cm} p{3.0cm}}
\textbf{TOA (BMJD)} & \textbf{Flux density (Jy)} & \textbf{Pulse width (ms)} & \textbf{Fluence (Jy\,ms)} \\ \hline
57769.6881141561(4) & 0.7(1)  & 3.56(9) & 2.6(4)  \\
57769.702301263(2)  & 0.17(3) & 3.4(4)  & 0.6(1)  \\
57769.7639680761(3) & 0.8(1)  & 2.05(6) & 1.7(3)  \\
57772.6649755688(7) & 0.36(5) & 2.1(1)  & 0.7(1)  \\
57772.688014045(3)  & 0.11(2) & 3.2(7)  & 0.35(9) \\
57772.758495566(1)  & 0.25(4) & 5.1(3)  & 1.3(2)  \\
57772.762396326(1)  & 0.22(3) & 1.8(2)  & 0.40(7) \\
57772.784720292(1)  & 0.21(3) & 1.9(3)  & 0.40(9) \\
57778.6885027615(7) & 0.39(6) & 2.6(1)  & 1.0(2)  \\
57778.756270766(2)  & 0.11(2) & 2.1(4)  & 0.23(6) \\
57778.756271161(2)  & 0.18(3) & 2.9(3)  & 0.5(1)  \\
57778.799193770(1)  & 0.21(3) & 1.8(2)  & 0.38(7) \\
57803.692917989(2)  & 0.21(3) & 3.5(3)  & 0.7(1)  \\
 \hline  
\end{tabular}
\end{minipage}
\end{table*}

Using the SDSS magnitude for the comparison star ($i'=17.25$) provided by Pan-STARRS\footnote{\url{http://archive.stsci.edu/panstarrs/search.php}} \citep{chambers16}, we calculated a minimum detectable optical burst flux density (mJy) by comparing the value of the noise in the FRB aperture (measured from the standard deviation in the light curve), with the mean counts measured from the comparison star in a single image. We extracted the comparison star counts in the same way as for the FRB location, conducting fixed aperture photometry and subtracting the sky contribution. The extracted counts for the comparison star were then calibrated into flux density units (mJy) using the Pan-STARRS magnitude and the standard AB conversion factor for a magnitude zero source of 3631 Jy \citep{fukugita96}. For the two sets of observations conducted in the SDSS $i'$-band (2017-02-16, 2017-02-19), we found 5-$\sigma$ detection limits of 1.9 mJy and 1.7 mJy respectively. 

To estimate the equivalent limit for the $i'\!+\!z'$-band observations (2017-01-16, 2017-01-19, 2017-01-25), we need to consider the difference in throughput between the $i'\!+\!z'$ and $i'$ filters. Using our existing model for the throughput of the TNT+ULTRASPEC configuration \citep{dhillon14,hardy17}, we find that a flat spectrum source should result in 1.5 times more detected photons in the $i'\!+\!z'$ filter compared to the $i'$ filter. We approximate the spectrum of the noise to be flat. We thus estimate that our $i'\!+\!z'$ burst limit is 1.5 times deeper than our $i'$ limit. Thus we obtain an $i'\!+\!z'$ band 5-$\sigma$ burst flux density detection limit of 1.2 mJy. 

For the stacked image, which consists of 20 $i'\!+\!z'$ frames and two $i'$ frames, we predict a 5-$\sigma$ detection limit of 0.33 mJy. This may not seem very deep, but it is actually the limit in the burst fluence that is most constraining, and most important. Since we predict that any optical component of the burst will be emitted at some point during the two nearest ULTRASPEC frames to the time of arrival of the radio burst, we can find the limit in the detectable fluence by multiplying the flux density limit of 0.33 mJy by the duration of the two nearest frames, which is 141.4~ms. The maximum simultaneous optical fluence for these bursts is therefore 0.046~Jy~ms. 

The only possible method for improving upon this limit would be to use a larger telescope and/or run at a faster cadence. Contrary to usual practice in astronomy, taking a deeper exposure does not help us. Had we taken a single 19.6-hour exposure with ULTRASPEC in grey time, we might expect to reach a persistent limiting magnitude of $i'=24.7$ at 5-$\sigma$, but this would represent a fluence limit for a single burst of 34~Jy~ms - a factor of 700 less constraining than our limit. This is because by scaling up the exposure time by a factor of 500\,000, we would decrease the effective signal-to-noise ratio by a factor of $\sqrt{500000}\simeq700$ due to the increased noise in the sky background (assuming no additional noise contribution from dark current, clock-induced charge or readout noise). In other words, short optical bursts would need to be much brighter to be detected in a long-exposure image, in order to overcome the noise contribution from the sky background. 

We can directly compare our optical fluence detection limit (at 767~nm) to the median radio fluence of the bursts observed by Effelsberg, which is 0.6 Jy~ms at 1.36 GHz. If emission in the radio and optical frequencies had the same intrinsic source, this would limit the broadband (optical to radio) spectral slope to being steeper than $f_{\nu} \propto \nu^{-0.2}$. This is roughly the same as the optical-to-radio spectral slope of the pulsed emission from the Crab pulsar \citep{dhillon11}, but we do not suggest a physical link here since there is currently no evidence that the emission in the radio and any undetected emission in the optical might be generated by the same mechanism. In fact, given that recent results find a break in the radio spectrum of FRB\,121102 \citep{law17}, we would not expect to see a continuous power law dependency from radio through to optical. Nevertheless, we can say that the optical fluence is at least 13 times lower than the radio fluence, giving an `optical-to-radio' spectral index of less than -0.2. At this stage we cannot rule out any suggested progenitor sources. 

\section{Periodicity}\label{sec:period}

\begin{figure}
 \hspace{-0.55cm}\includegraphics[width=.534\textwidth]{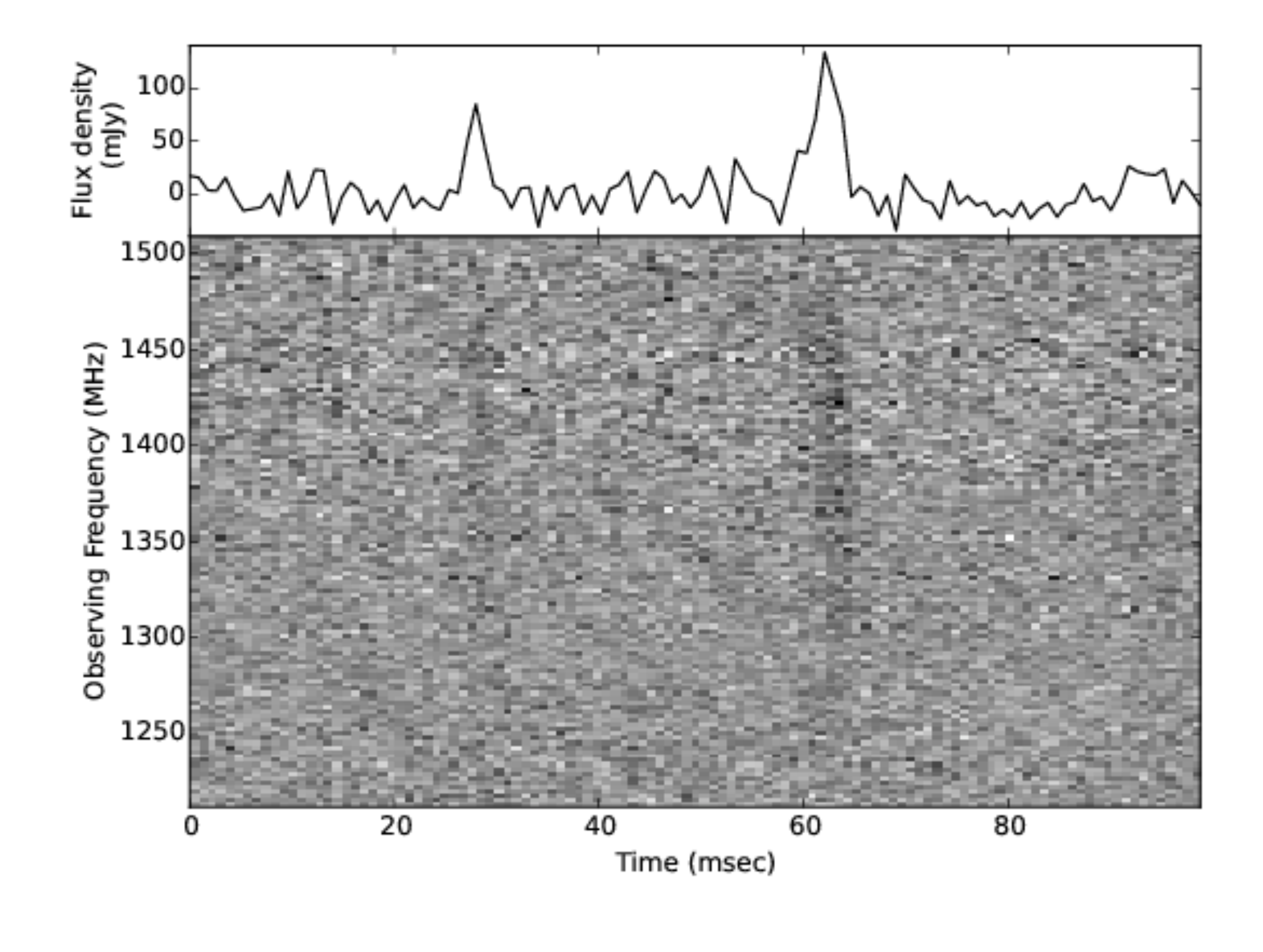}
 \vspace{-0.8cm}\caption{Profiles of the two bursts separated by only 34~ms, as observed by Effelsberg. The central greyscale panels show the total intensity versus observing frequency and time. Data have been downsampled (binned) in time by a factor of 16 and in frequency by factor of 4 to increase the S/N of the bursts. The upper sub-panels are burst profiles summed over all frequencies. The x-axis shows the topocentric time since 2017-01-25 18:02:07.59.}
 \label{fig:doubleburst}
\end{figure}

Two of the radio bursts we observed with Effelsberg have a separation of $\sim 34$~ms. These are clearly not overlapping in time, like those reported in \citet{champion15} and \citet{scholz16}, and so we do not consider them to be a single double-peaked burst. Figure \ref{fig:doubleburst} shows the burst profiles as a function of time. The two bursts are each consistent with Gaussian profiles with FWHM widths of $\sim 2$ and $\sim 3$~ms, respectively, and the signal level seen between the bursts is consistent with the noise floor. 

This separation may be an integer times an underlying spin periodicity, i.e. the period would be $\sim 34/N$~ms, with $N$ being some positive integer. However, we caution that this separation must not necessarily be an upper limit to an underlying periodicity; it could alternatively be a lower limit on the window in rotational phase during which radio bursts are emitted. A preliminary analysis of all the burst arrival times (those reported here and previously in the literature) did not yield a unique periodicity. 

Recently, \citet{scholz17} reported two bursts detected with the Green Bank Telescope which were separated by only 37~ms. Interpretation of this short separation alongside the 34~ms separation we observed is subject to the same caveats as stated above. A periodicity may be revealed but will require a more sophisticated analysis that considers, for example, multiple bursts detected during a single rotation in one or more rotational phase windows. Many Galactic radio-emitting neutron stars emit bursts in multiple rotational phase windows, as seen in the Crab pulsar \citep{hankins07}, in the rotating radio transients \citep{keane11a}, and in radio emission from Galactic magnetars \citep{serylak09,levin12}. However, measuring such a short underlying period with so few observed bursts is likely to be impossible, because after only a few cycles the errors in any individual arrival times will multiply up to something similar to the possible period itself, quickly leading to cycle number ambiguities. Furthermore, a periodicity analysis that includes bursts detected in multiple observing epochs will require not only searching in rotational period, but also period derivative (i.e. spin down) or orbital motion in the event that the source is in a binary.


\section{Conclusions}\label{sec:conclusion}
We have utilised the fast read-out capabilities of ULTRASPEC to search for optical burst counterparts to the repeating fast radio burst FRB\,121102. Our observations, conducted simultaneously with the Effelsberg 100-m radio telescope, provide a 5-$\sigma$ upper limit to the optical burst fluence of $0.046$~Jy~ms.  

We are only able to place weak limits on the broadband spectral index, in that it is steeper than $\alpha=-0.2$. Deeper searches for optical bursts would provide further constraints to this index, but this would only be possible with a larger telescope and an instrument that can run at a higher cadence. The upcoming HiPERCAM instrument, when mounted on the 10.4-m Gran Telescopio Canarias \citep{dhillon16}, will be capable of reaching a fluence detection limit of $0.0004$~Jy~ms, approximately 115 times deeper than the limit presented in this work, and 1500 times fainter than the median radio burst fluence. 

Two of our observed radio pulses are separated by only 34~ms. Whilst this may constrain possible searches for periodicity in the arrival times of these bursts, we caution that it does not rule out emission in multiple rotational phase windows of a longer period. 


\section*{Acknowledgements}
We thank the editor and anonymous referee for providing helpful comments that have improved this paper. We thank David Champion for helping with the Effelsberg observations, and Alex Kraus for assistance with scheduling.

VSD, SPL, ULTRASPEC and TRM are supported by STFC consolidated grants ST/J001589/1 and ST/L000733. LGS gratefully acknowledges financial support by the European Research Council for the ERC Starting Grant BEACON under contract no. 279702 and the Max Planck Society. RPA acknowledges funding from the European Research Council under the European Union's Seventh Framework Programme (FP/2007-2013) / ERC Grant Agreement no. 320964 (WDTracer). This work was based on observations with the 100-m telescope of the MPIfR (Max-Planck-Institut f\"{u}r Radioastronomie) at Effelsberg, Germany, and on observations made with the Thai National Telescope, operated by the National Astronomical Research Institute of Thailand. We acknowledge use of NASA's Astrophysics Data System Bibliographic Services, and the Pan-STARRS database.

\bibliographystyle{mn2e}
\bibliography{abbrev.bib,refs.bib}


\label{lastpage}

\end{document}